\documentclass{article}
\usepackage{spconf,amsmath,graphicx,booktabs}
\usepackage[font=small]{caption}

\title{cloning one's voice using very limited data in the wild}
\name{\begin{tabular}{c}Dongyang Dai$^{*}${ \thanks{$^{*}$ daidongyang@bytedance.com}}, Yuanzhe Chen\thanks{{Demos on https://daidongyang.github.io/hierademo/index.html}}, Li Chen,  Ming Tu, Lu Liu \\Rui Xia, Qiao Tian, Yuping Wang, Yuxuan Wang\end{tabular}}
\address{Speech, Audio \& Music Intelligence (SAMI), ByteDance Inc.}
\begin{document}
%
\maketitle
\begin{abstract}
With the increasing popularity of speech synthesis products, the industry has put forward more requirements for personalized speech synthesis: (1) How to use low-resource, easily accessible data to clone a person’s voice. (2) How to clone a person’s voice while controlling the style and prosody. To solve the above two problems, we proposed the Hieratron model framework in which the prosody and timbre are modeled separately using two modules, therefore, the independent control of timbre and the other characteristics of audio can be achieved while generating speech. The practice shows that, for very limited target speaker data in the wild, Hieratron has obvious advantages over the traditional method, in addition to controlling the style and language of the generated speech, the mean opinion score on speech quality of the generated speech has also been improved by more than 0.2 points.
\end{abstract}
\section{Introduction}
As an important part of human-computer interaction, speech synthesis has attracted the attention of researchers for decades. With the advent of end-to-end acoustic models such as Tacotron\cite{wang2017tacotron}, Fastspeech\cite{ren2019fastspeech} and neural network-based vocoders such as WaveNet\cite{oord2016wavenet} and Mel-GAN\cite{kumar2019melgan}, the naturalness and audio quality of the generated speech have been greatly improved, and speech synthesis technology has appeared in more and more products. Furthermore, the industry has an increasing demand for personalized speech synthesis, how to use fewer resources and build a robust and controllable personalized speech synthesis system has become a current research hotspot.

To clone a person’s voice from a limited resource, Jia et al. \cite{jia2018transfer} combined a speaker encoder, which was pre-trained on the speaker verification task, and Tacotron2\cite{shen2018natural} into a multi-speaker speech synthesis model. The speaker encoder model takes reference waveform as input and computes a fixed-length vector representing speaker information as the conditional input of Tacotron2. During training, the target waveform of the total model (ground truth) is used as the reference waveform.When inferring, the reference waveform does not have to be consistent with the input text content, which is any sentence from the target speaker's voice, and the timbre of the synthesized waveform is similar to the reference waveform. Jia's approach \cite{jia2018transfer} supports zero-shot learning, however, for out-of-set speaker, the Tacotron2 model often introduces bad cases such as pause errors and noise for speaker representations that have not been seen before.

In order to obtain a voice cloning system with higher speaker similarity and more stable synthesis results, Arik et al.\cite{arik2018neural} proposed a strategy of pre-training a multi-speaker TTS model first, and then fine-tuning on a specific speaker data. In \cite{arik2018neural}, speaker embedding, which is trainable or from a pre-trained speaker verification model, will be conditioned to the end2end speech synthesis model. This method needs to accumulate a large amount of high-quality labeled multi-speaker TTS data for pre-training, which is very costly. For the new target speaker data, in addition to the requirement of  being clean and free of noise, the speaker also needs to have accurate pronunciation and relatively fluent expression in order to synthesize stable results.

Voice conversion is a technique to modify the speech from a source speaker to make it sound similar to a target speaker while keeping the linguistic content unchanged. Sun et al. \cite{sun2016phonetic} proposed to use a speaker-independent speech recognition model to extract phonetic posteriorgrams (PPGs) from source speech (uttered by the source speaker) , and then use the target speaker data to map PPGs to the target speaker’s speech. Cao et al. \cite{cao2020code} used PPGs in cross-language speech synthesis tasks, and the synthesized speech has been improved in terms of speech intelligibility and audio fidelity.  In \cite{wang2020spoken}, the parameter space of the original speaker's voice was designed to be decomposed into two subspaces, one for modeling spoken content and the other one for the speaker 's voice. In addition, finetuning on the two subspaces can effectively reduce the overfitting issue.

In this paper, we proposed a hierarchical speech synthesis model framework named as Hieratron.  With the help of bottleneck features extracted from a pre-trained automatic speech recognition (ASR) model, the speech synthesis acoustic model is separated into two different modules: the \textit{text2bottleneck} module and the \textit{bottleneck2mel} module. The \textit{bottleneck2mel} is for modeling timbre information, and the \textit{text2bottleneck} module is for modeling content, prosody, and other information. These two modules can be trained separately using different sources of data: the \textit{text2bottleneck} module uses high-quality and well-annotated TTS data, and the \textit{bottleneck2mel} can leverage easy-to-collect massive multi-speaker data for training. Compared to the previous voice cloning framework like  \cite{jia2018transfer} and \cite{arik2018neural}, our approach has the following three advantages:
\begin{enumerate}
	\item The independent control of timbre and prosody: In our approach, the prosody and timbre are modeled by different modules. Therefore, when synthesizing speech, different prosody and timbre can be combined arbitrarily to achieve interesting synthesis effects, such as cross style voice cloning.
	\item Less requirement for the target speaker data's quality: Compared to the previous method which requires accurate pronunciation and fluent expression data for the target speaker, the \textit{bottlenek2mel} module, proposed in this paper, can robustly model timbre of the new speaker with wild data.
	\item Less requirement for the target speaker data's amount: For the voice cloning task, the more speakers used in pre-training, the less new speaker data is needed for fine-tuning. In our method, the bottleneck2mel module barely needs any data annotation, Therefore, the cost of collecting data is low. It's easy to collect thousands of speaker data to pre-train the \textit{bottleneck2mel} module, which makes it possible to perform stable synthesis results after fine-tuning on very little target speaker data.
\end{enumerate}

In addition, we creatively proposed a ASR-bottleneck-based \textit{prosody classifier}, without the requirement of any prosody annotation information. This classifier takes the timbre-independent ASR bottleneck as input, and the speaker id as classification label. We assumed that different speakers are distinguished by timbre and prosody, and the input bottleneck does not contain timbre information, so the classifier will distinguish different speakers according to prosody. When cloning using low-resource low-quality new speaker data, for stable synthesis result, we only fine-tune the \textit{bottleneck2mel} module while frozening the \textit{text2bottleneck} module. To maintain the prosody similarity between the target speaker data (training data) and the synthesized speech, we train multiple \textit{text2bottleneck} models with different prosodies on multi-speaker TTS data, then during inference, the prosody classifier is used to find the corresponding \textit{text2bottleneck} model whose prosody best matches the target speaker.

\section{The proposed approach}
This part is divided into three subsections. Section~\ref{subsec21} introduces the Hieratron system as a whole, Section~\ref{subsec22} introduces some details that have a great impact on the quality of generated speech, and Section~\ref{subsec23} further analyzes the advantages and limitations of the Hieratron framework.
\subsection{The  Hieratron system}
\label{subsec21}
\begin{figure}[t]
	\vspace{-10pt}
	\centering
	\includegraphics[scale=0.42]{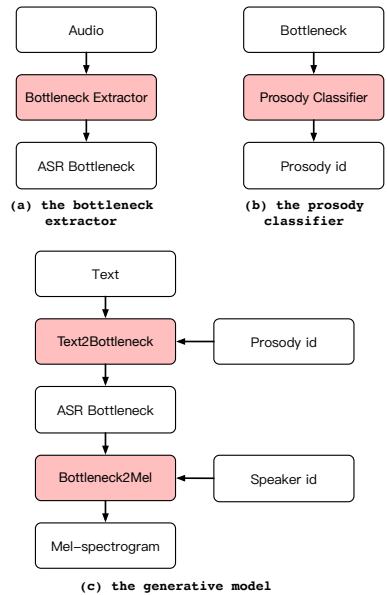}
	\vspace{-5pt}
	\caption{The Hieratron system}
	\label{fig:Hieratron-systeml}
	\vspace{-15pt}
\end{figure}
Shown in Fig.~\ref{fig:Hieratron-systeml}, Hieratron system consists of 4 sub-modules, namely \textit{bottleneck extractor}, \textit{prosody classifier}, \textit{text2bottleneck} and \textit{bottleneck2mel}. 

The  \textit{bottleneck extractor} is part of a pre-trained speaker-independent automatic speech recognition model (SI-ASR), which accepts voice waveform as input and outputs bottleneck irrelevant to the speaker's timbre.

The \textit{prosody classifier} is trained on high-quality multi-speaker TTS data, which accepts ASR bottleneck extracted by \textit{bottleneck extractor} as input, and outputs speaker id. As ASR bottleneck is irrelevant to the speaker timbre, the classifier relies on prosody to distinguish different speakers, we marked speaker id as prosody id.

The \textit{text2bottleneck} is trained also on high-quality multi-speaker TTS data, the same as \textit{prosody classifier}. As multi-speaker TTS data has format as \textit{((text, speaker id), wav)}, we marked speaker id as prosody id and with the help of \textit{bottleneck extractor}, we can get \textit{text2bottleneck}'s training data with the format as \textit{((text, prosody id), ASR bottleneck)}, the \textit{text2bottleneck} module establishes the mapping relationship between \textit{(text, prosody id)} and ASR bottleneck.

The \textit{bottleneck2mel} is pre-trained on multi-speaker data, and fine-tuned on the target speaker data. Except for speaker id, training data does not require any other annotations, so it is easy to collect data with thousands of speakers to pre-train the \textit{bottleneck2mel} model. The original format of training data is \textit{(speaker id, wav)}, with the the help \textit{bottleneck extractor} and digital signal processing, we can get ASR bottleneck and Mel-spectrogram corresponding to original wav, so format of the training data could be \textit{((ASR bottleneck, speaker id ), Mel-spectrogram)}.  the \textit{bottleneck2mel} module establishes the mapping relationship from \textit{(ASR bottleneck, speaker id)} to Mel-spectrogram.

The \textit{prosody classifier} and \textit{text2bottleneck} is trained on high-quality well-labeled multi-speaker TTS data, and \textit{bottleneck2mel} is pre-trained on speech data composed of thousands of speakers. When encountering a new speaker data in the wild, to synthesize to the new speaker's voice, the new speaker data will be used in two ways: (1) the new speaker data will be fed to the prosody classifier module to find prosody most similar to the new speaker in the multi-speaker TTS data, (2) the new speaker data will be used to fine-tune the \textit{bottleneck2mel}. When inferencing, the \textit{text2bottleneck} accepts text and prosody id corresponding to most similar prosody found in the multi-speaker TTS data as input, and outputs the predicted ASR bottleneck, then the fine-tuned \textit{bottleneck2mel} takes the predicted ASR bottleneck as input and outputs corresponding Mel-spectrogram, finally, the universal neural-vocoder could convert the Mel-spectrogram to the waveform whose timbre is the same as the new speaker.

Optionally, the \textit{text2bottleneck} module can also be trained on the TTS data of a single speaker. In this case, the \textit{prosody classifier} module is no longer needed, and the synthesized speech has only one kind of prosody Hieratron without prosody selection (PS) is meaningful for some scenarios, such as limited TTS data, cross dialect voice cloning, cross style voice cloning. In addition, to emphasize the prosodic characteristics of the \textit{text2bottleneck}, we can predict also the normalized fundamental frequency and energy together with the ASR bottleneck by the \textit{text2bottleneck}.

\subsection{Noteworthy details}
\label{subsec22}
The \textit{bottleneck extractor} is part of SI-ASR model, the \textit{prosidy classifier}  module is implemented by CNN + RNN, and we used modified Tacotron2 or Fastspeech as \textit{text2bottleneck} module, the \textit{bottleneck2mel} module is implemented by stacked self-attention layers.

The ASR bottleneck, extracted by SI-ASR model\footnote{For Chinese, we applied the encoder of conformer\cite{gulati2020conformer} to extract bottleneck, and for other languages, we applied fine-tuned wav2vec2\cite{baevski2020wav2vec}.} model, has a decisive influence on the quality of synthesized speech. When training the SI-ASR model, certain restrictions are conducive to better training and convergence of the \textit{text2bottleneck}, whose performance is the limit of the entire Hieratron system. According to the author’s experience, (1) The generated speech could be more natural and expressive using bottleneck with high time resolution than low time resolution. (2) The ASR model that uses phoneme as label is more conducive to the reduction of synthesized speech's pronunciation errors than using character or byte pair encoding (bpe) as label. (3) When training ASR model, normalizing the bottleneck and limiting its range, such as [-4, 4], are helpful for training the \textit{text2bottleneck} more stably, reducing pronunciation errors and other badcases.

\begin{figure}[t]
	\vspace{-10pt}
	\centering
	\includegraphics[scale=0.42]{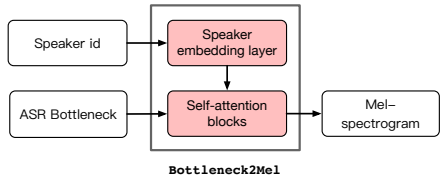}
	\vspace{-5pt}
	\caption{The bottleneck2mel model}
	\label{fig:bottleneck2mell}
	\vspace{-15pt}
\end{figure}

The \textit{bottleneck2mel} module takes ASR bottleneck and speaker id as input, and outputs the Mel-spectrogram.  As shown in Fig.~\ref{fig:bottleneck2mell}, the speaker id is in the form of one-hot, and passed through the speaker embedding layer before being conditioned to the self-attention blocks.  When fine-tuning the \textit{bottleneck2mel} on the new speaker data, for better performance on a very small amount of data, we applied the following two tricks: (1) For the new speaker's speech, we first used the voice activity detection (VAD) tool to divide these voices into many small segments, then shuffle these small segments and randomly combine them to form new audios multiple times to achieve data augmentation. (2) When fine-tuning, we first frozen the parameters of the self-attention blocks, and only update the parameters of the speaker embedding layer. After a few thousand steps, update the parameters of the entire \textit{bottleneck2mel} (speaker embedding layer + self-attention blocks).

For cross-language voice cloning, with the speech of the target speaker speaking language A (source language) only, we need to synthesize the speech target speaker speaking language B (target language). We applied the target language's data to train SI-ASR model, which is more suitable for training the \textit{text2bottleneck}. It has been verified that the bottleneck extracted by wav2vec2\cite{baevski2020wav2vec} fine-tuned on 500 hours' data are on par with the bottleneck extracted by the conformer\cite{gulati2020conformer} training on 7000 hours for our task. 

\subsection{Further discussion}
\label{subsec23}
The \textit{text2bottleneck} is trained on high-quality TTS data, and the \textit{bottleneck2mel} is trained on easy collected data without any labeling except for speaker id. The \textit{bottleneck2mel} module is used for only modeling the timbre, so even if the new (target) speaker's voice is not fluent or expressive enough, the pronunciation is not standard, high-quality speech with the target speaker's timbre could also be synthesized in Hieratron framework. In theory, the upper limit of synthesized speech's quality is the quality of the high-quality TTS data used to train \textit{text2bottleneck}, that is to say, the quality of synthesized speech is hopeful to exceed the original target speaker data in Hieratron framework.

For Hieratron, the ideal bottleneck should be a speech representation that does not contain any timbre information at all without losing any other information. However, what information the bottleneck contains is not under our complete control as it is extracted by the ASR system. This is also an important factor that affects the audio quality of synthesized speech, in fact, for the case where the quality of the target speaker is quite high, the end-to-end \textit{text2mel} model will have better audio quality than the Hieratron solution.

In Hieratron framework, it is feasible to use the target speaker's recording of about one minute to perform stable voice cloning. As the bottleneck extracted by the ASR system does not contain noise information, for the case where the target speaker's speech data contains noise, we applied a neural network-based speech enhancement model to denoise before training on the target speaker data. Combined with the above analysis of audio quality, the authors believe that it is meaningful to explore bottleneck extraction methods beyond the ASR system.

\section{Subjective evaluation and analysis}
To indicate the advantages and other characteristics of Hieratron, here we show some subjective evaluation results.
\subsection{Compared with  text2mel model}

We compared the speech generated by the \textit{text2mel} method in \cite{arik2018neural} and Hieratron without prosody selection, in the scene of amateurs recording. The \textit{text2mel} and the Hieratron both use data of 1700 speakers to pre-train the average model, and the pre-trained models are fine-tuned on phone recordings of four target speakers (where spk-1, spk-2 are females, spk-3, spk-4 are males, the recording time of each speaker is 2 minutes). The synthesized speech is evaluated on speech quality and speaker similarity separately\footnote{ Out of cost considerations, a male and a female were randomly selected from the four speakers for speaker similarity evaluation, refer to the appendix for details on subjective evaluation}.

\begin{table} [t]
	\caption{The MOS result in term of, (a) speech quality, (b) speaker similarity}
		\vspace{-10pt}
	\label{table:mos1}
	\begin{minipage}[b]{.48\linewidth}
		
		\centerline{
			\scalebox{0.7}{
				\begin{tabular}{c|cccc}
					\hline
					& text2mel & without PS \\
					\hline
					\textbf{spk-1}       &       3.47       &   3.78   \\
					\textbf{spk-2}      &      3.38       &     3.64       \\
					\textbf{spk-3}       &        2.90      &    3.05    \\
					\textbf{spk-4}        &       2.79       &        3.11      \\
					\hline
		\end{tabular}}}
		\centerline{\small(a) }\medskip
	\end{minipage}
	\hfill
	\begin{minipage}[b]{0.48\linewidth}
		\centerline{
			\scalebox{0.7}{
				\begin{tabular}{c|cccc}
					\hline
					& text2mel & without PS  \\
					\hline
					\textbf{spk-1}       &       4.14       &    4.17    \\
					\textbf{spk-4}        &      4.67       &      4.57       \\
					\hline
		\end{tabular}}}
		\centerline{\small(b)}\medskip
	\end{minipage}
	\vspace{-15pt}
\end{table}

 \begin{table} [t]
	\caption{The MOS result in term of, (a) speech quality, (b) speaker similarity}
	\vspace{-10pt}
	\label{table:mos2}
	\begin{minipage}[b]{.48\linewidth}
		
		\centerline{
			\scalebox{0.7}{
				\begin{tabular}{c|cccc}
					\hline
					& without PS & with PS \\
					\hline
					\textbf{spk-a}       &       3.28       &    3.17    \\
					\textbf{spk-b}      &      3.32       &     2.89       \\
					\textbf{spk-c}       &        3.74      &   3.64    \\
					\textbf{spk-d}        &       3.71       &        3.61        \\
					\textbf{spk-e}        &        3.32       &     3.02     \\
					\textbf{spk-f}        &         3.81            &    3.79     \\
					\hline
		\end{tabular}}}
		\centerline{\small(a) }\medskip
	\end{minipage}
	\hfill
	\begin{minipage}[b]{0.48\linewidth}
		\centerline{
			\scalebox{0.7}{
				\begin{tabular}{c|cccc}
					\hline
					& without PS& with PS  \\
					\hline
					\textbf{spk-a}       &       3.29      &    3.25   \\
					\textbf{spk-b}      &      2.68       &     2.98       \\
					\textbf{spk-c}       &        2.79      &   2.87    \\
					\textbf{spk-d}        &       3.05      &        3.02        \\
					\textbf{spk-e}        &         2.80        &    2.72 \\
					\textbf{spk-f}        &          2.66   &   2.73 \\
					\hline
		\end{tabular}}}
		\centerline{\small(b)}\medskip
	\end{minipage}
	\vspace{-15pt}
\end{table}

The MOS from different subjective evaluation listeners is shown in Table-\ref{table:mos1}. From Table-\ref{table:mos1}, it can be seen that the speech quality of audio generated by Hieratron is 0.26 points higher than that of \textit{text2mel} on average, and the speaker similarity is reduced only by 0.035 points on average. That is to say, compared to \cite{arik2018neural}, in the scene of a very small amount of target speaker data in the wild, Hieratron framework can synthesize voice with a significantly higher degree of naturalness while the speaker similarity is reduced limited. 

%
%

 \subsection{The influence of prosody selection}
 
 Six different speakers were asked to record their voice by mobile phones or earphones of any content, and the recording duration they provide ranges from 30 seconds to 3 minutes. We have carried out different settings of Hieratron on these data, including Hieratron without prosody selection, Hieratron with prosody selection, cross style voice cloning, cross language voice cloning, etc.\footnote{Generated audios: https://daidongyang.github.io/hierademo/index.html} 
 
 For Hieratron with prosody selection, the \textit{text2bottleneck} is trained on 122 speakers' TTS data, representing 122 kinds of prosody, of which 16 speakers have more than 5,000 samples, and 106 speakers have only 500 samples. For Hieratron without prosody selection, only one of the 16 speaker's data which have more than 5000 samples is used to train the \textit{text2bottleneck}.  Subjective evaluation is conducted to compare and analyze the effect of prosody selection. And the MOS is shown in Table~\ref{table:mos2}.
 
 According to Table~\ref{table:mos2}, it can be found that after using prosody selection,  the speaker similarity of synthesized voice is improved by 0.05 points, while the speech quality is reduced by 0.18 points. According to the evaluation results of speaker similarity, we found that for most people, changing the speaker’s prosody does not have a great impact on identifying a speaker, except for those whose prosody are very distinctive, like spk-b. In addition, for the speech generated with prosody selection, the naturalness and quality of the synthesized speech under each target speaker is influenced by the data used to train \textit{text2bottleneck} which has the most similar prosody to the target speaker, since the data quality of the selected speaker with the most similar prosody is worse than the TTS data used for training Hieratron without prosody selection, this leads to a reduction in the score of speech quality, except for spk-f (the selected speaker's data closest to the spk-f prosody have also than 5000 samples). This shows that if there are enough high-quality TTS data of multiple speakers, the prosody selection scheme is feasible.
 
%
%

 \section{Conclusion}

In this paper, we proposed a voice cloning model framework named Hieratron, whose generative model consists of \textit{text2bottleneck} and \textit{bottleneck2mel} modules. The two modules are trained on different sources of data, therefore cloning one's voice on a small amount of low-quality target speaker data is realized,  furthermore, more control over the synthesized voice could also be achieved, such as cross language voice cloning, cross style voice cloning.
\bibliographystyle{IEEEbib}
\bibliography{strings,refs}

\begin{thebibliography}{10}

\bibitem{wang2017tacotron}
Yuxuan Wang, RJ~Skerry-Ryan, Daisy Stanton, Yonghui Wu, Ron~J Weiss, Navdeep
  Jaitly, Zongheng Yang, Ying Xiao, Zhifeng Chen, Samy Bengio, et~al.,
\newblock ``Tacotron: Towards end-to-end speech synthesis,''
\newblock {\em arXiv preprint arXiv:1703.10135}, 2017.

\bibitem{ren2019fastspeech}
Yi~Ren, Yangjun Ruan, Xu~Tan, Tao Qin, Sheng Zhao, Zhou Zhao, and Tie-Yan Liu,
\newblock ``Fastspeech: Fast, robust and controllable text to speech,''
\newblock {\em arXiv preprint arXiv:1905.09263}, 2019.

\bibitem{oord2016wavenet}
Aaron van~den Oord, Sander Dieleman, Heiga Zen, Karen Simonyan, Oriol Vinyals,
  Alex Graves, Nal Kalchbrenner, Andrew Senior, and Koray Kavukcuoglu,
\newblock ``Wavenet: A generative model for raw audio,''
\newblock {\em arXiv preprint arXiv:1609.03499}, 2016.

\bibitem{kumar2019melgan}
Kundan Kumar, Rithesh Kumar, Thibault de~Boissiere, Lucas Gestin, Wei~Zhen
  Teoh, Jose Sotelo, Alexandre de~Br{\'e}bisson, Yoshua Bengio, and Aaron
  Courville,
\newblock ``Melgan: Generative adversarial networks for conditional waveform
  synthesis,''
\newblock {\em arXiv preprint arXiv:1910.06711}, 2019.

\bibitem{jia2018transfer}
Ye~Jia, Yu~Zhang, Ron~J Weiss, Quan Wang, Jonathan Shen, Fei Ren, Zhifeng Chen,
  Patrick Nguyen, Ruoming Pang, Ignacio~Lopez Moreno, et~al.,
\newblock ``Transfer learning from speaker verification to multispeaker
  text-to-speech synthesis,''
\newblock {\em arXiv preprint arXiv:1806.04558}, 2018.

\bibitem{shen2018natural}
Jonathan Shen, Ruoming Pang, Ron~J Weiss, Mike Schuster, Navdeep Jaitly,
  Zongheng Yang, Zhifeng Chen, Yu~Zhang, Yuxuan Wang, Rj~Skerrv-Ryan, et~al.,
\newblock ``Natural tts synthesis by conditioning wavenet on mel spectrogram
  predictions,''
\newblock in {\em 2018 IEEE International Conference on Acoustics, Speech and
  Signal Processing (ICASSP)}. IEEE, 2018, pp. 4779--4783.

\bibitem{arik2018neural}
Sercan~O Arik, Jitong Chen, Kainan Peng, Wei Ping, and Yanqi Zhou,
\newblock ``Neural voice cloning with a few samples,''
\newblock {\em arXiv preprint arXiv:1802.06006}, 2018.

\bibitem{sun2016phonetic}
Lifa Sun, Kun Li, Hao Wang, Shiyin Kang, and Helen Meng,
\newblock ``Phonetic posteriorgrams for many-to-one voice conversion without
  parallel data training,''
\newblock in {\em 2016 IEEE International Conference on Multimedia and Expo
  (ICME)}. IEEE, 2016, pp. 1--6.

\bibitem{cao2020code}
Yuewen Cao, Songxiang Liu, Xixin Wu, Shiyin Kang, Peng Liu, Zhiyong Wu, Xunying
  Liu, Dan Su, Dong Yu, and Helen Meng,
\newblock ``Code-switched speech synthesis using bilingual phonetic
  posteriorgram with only monolingual corpora,''
\newblock in {\em ICASSP 2020-2020 IEEE International Conference on Acoustics,
  Speech and Signal Processing (ICASSP)}. IEEE, 2020, pp. 7619--7623.

\bibitem{wang2020spoken}
Tao Wang, Jianhua Tao, Ruibo Fu, Jiangyan Yi, Zhengqi Wen, and Rongxiu Zhong,
\newblock ``Spoken content and voice factorization for few-shot speaker
  adaptation.,''
\newblock in {\em INTERSPEECH}, 2020, pp. 796--800.

\bibitem{gulati2020conformer}
Anmol Gulati, James Qin, Chung-Cheng Chiu, Niki Parmar, Yu~Zhang, Jiahui Yu,
  Wei Han, Shibo Wang, Zhengdong Zhang, Yonghui Wu, et~al.,
\newblock ``Conformer: Convolution-augmented transformer for speech
  recognition,''
\newblock {\em arXiv preprint arXiv:2005.08100}, 2020.

\bibitem{baevski2020wav2vec}
Alexei Baevski, Henry Zhou, Abdelrahman Mohamed, and Michael Auli,
\newblock ``wav2vec 2.0: A framework for self-supervised learning of speech
  representations,''
\newblock {\em arXiv preprint arXiv:2006.11477}, 2020.

\end{thebibliography}

\appendix

\section{The Subjective evaluation}
\small{In the evaluation, each set of experimental configurations for each speaker is 30 synthesized speeches evaluated by 10 different listeners, and each person gives the evaluation results separately. The subjective evaluation of speech quality and speaker similarity adopts a 5-point system, and the evaluation score can only be an integer between 1-5. The meaning of each score is as follows:}
\vspace{-5pt}
\subsection{Speech quality}
\small{
	1: Can’t understand, can only understand a few words \\
	2: Some key words are unclear, and the pauses and pronunciation make people feel uncomfortable \\
	3: Generally understandable and acceptable. The rhythmic pause is not good enough\\
	4: Natural, clear and understandable, good hearing, willing to accept \\
	5: Broadcasting level, unable to distinguish between human voice and synthesized voice
}
\vspace{-5pt}
\subsection{Speech similiarity}
\subsubsection{Compared with tacotron based text2mel model}
\small{
	1: This is definitely not the same person, even the gender is different. \\
	2: It's not like the same person, most of them don't resemble, there are individual details.\\
	3: It may be the same person, but I can't be sure. \\
	4: Should be the same person, although there are some differences \\
	5: This must be the same person, the tone and the way of speaking are the same person
}

\subsubsection{The influence of prosody selection} 
\small{
	1: This is definitely not the same person, even the gender is different. \\
	2: It may be the same person, but I can't be sure.\footnote{To highlight the role of prosody selection, speaker similarity here uses more stringent standards} \\
	3: Should be the same person, although there are some differences\\
	4: High probability be the same person. Not only heard as one person, but the tone and intonation are also somewhat similar. \\
	5: Must be the same person, their voice, intonation and speaking style are very similar.
}

\end{document}